\documentclass[aps,pra,onecolumn,groupedaddress,showpacs,nofootinbib]{revtex4-1}
\usepackage{graphicx}
\usepackage{amsmath}
\usepackage{bbm}
\usepackage[bbgreekl]{mathbbol}
\DeclareMathSymbol\bbDelta  \mathord{bbold}{"01}% 
\begin{document}

%%%%%%%%%%%%%%%%%%%%%%%%%%%%%%%%%%%%%%
%      
%%%%%%%%%%%%%%%%%%%%%%%%%%%%%%%%%%%%%%
%%%%%%%%%%%%%%%%%%%%%%%%%%%%%%%%%%%%%%
%       TITLE, AUTHOR(S) AND ABSTRACT
%%%%%%%%%%%%%%%%%%%%%%%%%%%%%%%%%%%%%%
%%%%%%%%%%%%%%%%%%%%%%%%%%%%%%%%%%%%%%
%       
%%%%%%%%%%%%%%%%%%%%%%%%%%%%%%%%%%%%%%

\title{Theory of Rayleigh-Brillouin optical activity light scattering applicable to chiral liquids}

\author{Robert P. Cameron}
\email{robert.p.cameron@strath.ac.uk}
\address{SUPA and Department of Physics, University of Strathclyde, Glasgow G4 0NG, United Kingdom}

\author{Emmanouil I. Alexakis}
\address{SUPA and Department of Physics, University of Strathclyde, Glasgow G4 0NG, United Kingdom}

\author{Aidan S. Arnold}
\address{SUPA and Department of Physics, University of Strathclyde, Glasgow G4 0NG, United Kingdom}

\author{ Duncan McArthur}
\address{SUPA and Department of Physics, University of Strathclyde, Glasgow G4 0NG, United Kingdom}

\begin{abstract}
It has long been understood that dilute samples of chiral molecules such as rarefied gases should exhibit Rayleigh optical activity. We extend the existing theory by accounting for molecular dynamics and correlations, thus obtaining a more general theory of Rayleigh-Brillouin optical activity applicable to dense samples such as neat liquids.
\end{abstract}

\maketitle

%%%%%%%%%%%%%%%%%%%%%%%%%%%%%%%%%%%%%%
%      
%%%%%%%%%%%%%%%%%%%%%%%%%%%%%%%%%%%%%%
%%%%%%%%%%%%%%%%%%%%%%%%%%%%%%%%%%%%%%
%       INTRODUCTION
%%%%%%%%%%%%%%%%%%%%%%%%%%%%%%%%%%%%%%
%%%%%%%%%%%%%%%%%%%%%%%%%%%%%%%%%%%%%%
%       
%%%%%%%%%%%%%%%%%%%%%%%%%%%%%%%%%%%%%%

\section{Introduction}
\label{Introduction}
It was predicted a little over fifty years ago by Barron and collaborators that chiral molecules should exhibit Rayleigh optical activity (RayOA): differential Rayleigh scattering with respect to left- and right-handed circular polarisations of light \cite{Atkins69a, Barron71a, Barron04a}. In the theoretical descriptions of RayOA published to date, each molecule is effectively held static in position and orientation and scattered intensities due to different molecules are added incoherently \cite{Atkins69a, Barron71a, Andrews80a, Hecht94a, Craig98a, Barron04a, Zuber08a, Cameron14a, Cameron18a, Forbes19a, Forbes19b}. These descriptions provide expressions for the total intensity of the analysed signal and are best suited to dilute samples such as rarefied gases in which correlations between molecules are unimportant \cite{Landau84a, Berne00a, Sobelman02a, Boyd03a, Barron04a}.

In this paper, we extend the existing theory of RayOA by accounting for the translational and rotational dynamics of the molecules and adding scattered fields due to different molecules coherently, thus obtaining a more general theory that describes Rayleigh-Brillouin optical activity (RayBOA). Our theory provides expressions for not only the total intensity of the analysed signal but also the underlying frequency spectrum. It is applicable to dilute samples such as rarefied gases as well as dense samples such as neat liquids in which correlations between molecules are important.

Raman optical activity (ROA) \cite{Atkins69a, Barron71a, Barron73a, Hug75a, Barron04a, Pour18a} is the inelastic sister of RayOA. The theory of ROA does not need to be extended like the theory of RayOA, however, as one can add Raman scattered intensities due to different molecules incoherently at essentially all sample densities \cite{Barron04a, Michal22a}. 

(Linear) RayOA is not to be confused with hyper Rayleigh optical activity (HRS OA) \cite{Andrews79a, Collins19a, Verreault20a, Ohnoutek20a, Ohnoutek22b} and its extensions \cite{Ohnoutek21a, Ohnoutek22a}, which are distinct, nonlinear optical phenomena. 

For large biological scatterers, the  terminology ``circular intensity differential scattering'' (CIDS) is often used \cite{Maestre82a, Tinoco84a, Gratiet20a, Pan22a, Gasso22a} for what is essentially RayOA.

%%%%%%%%%%%%%%%%%%%%%%%%%%%%%%%%%%%%%%
%      
%%%%%%%%%%%%%%%%%%%%%%%%%%%%%%%%%%%%%%
%%%%%%%%%%%%%%%%%%%%%%%%%%%%%%%%%%%%%%
%       THEORY OF RAYBOA
%%%%%%%%%%%%%%%%%%%%%%%%%%%%%%%%%%%%%%
%%%%%%%%%%%%%%%%%%%%%%%%%%%%%%%%%%%%%%
%       
%%%%%%%%%%%%%%%%%%%%%%%%%%%%%%%%%%%%%%

\section{Theory of RayBOA}
\label{Theory of RayBOA}
\begin{figure}[h!]
\centering
\includegraphics[width=0.5\linewidth]{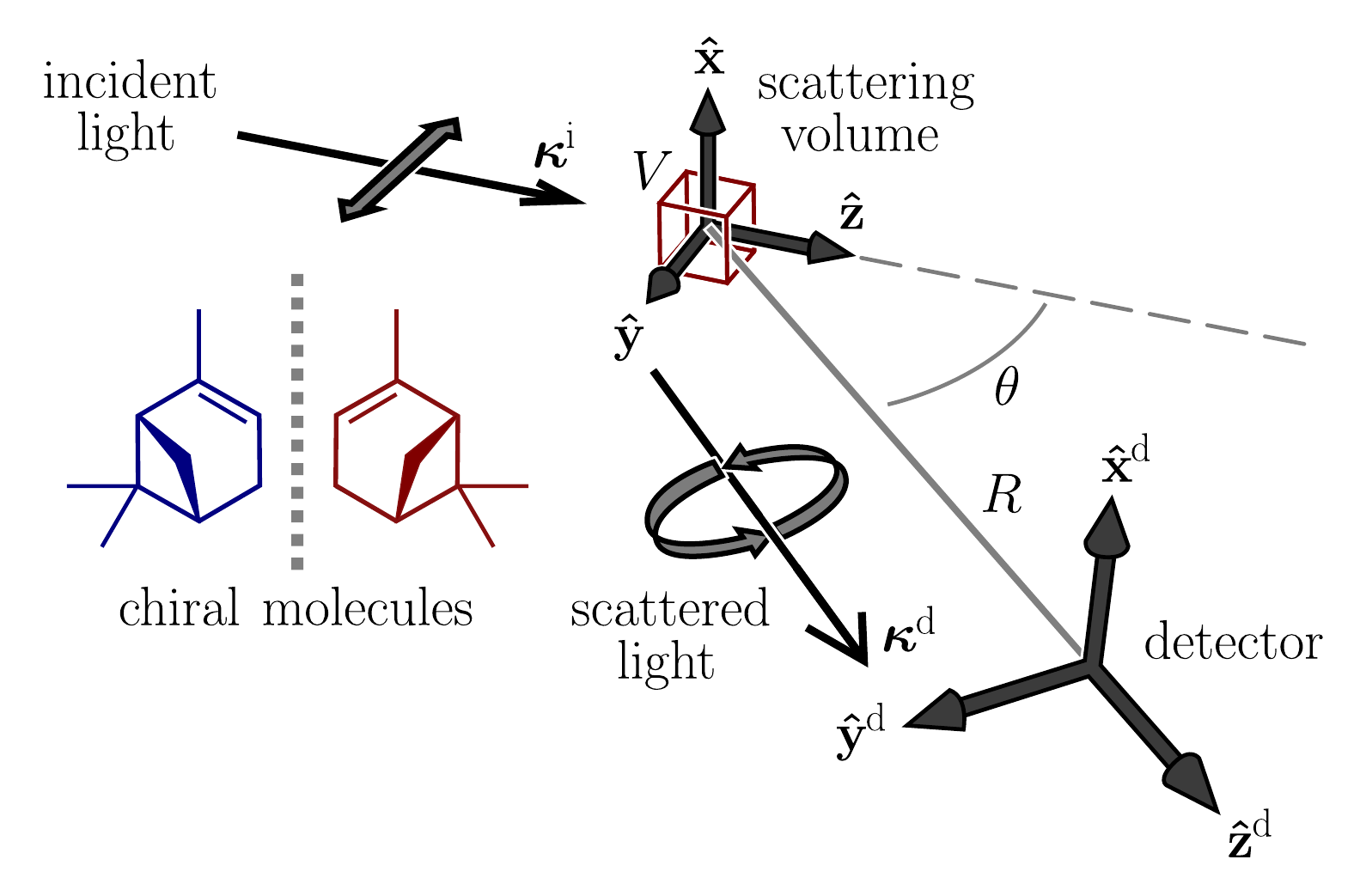}
\caption{\small Our scattering geometry, illustrated for neat ($1R$,$5R$)-$\alpha$-pinene and an SCP configuration.}
\label{Fig1}
\end{figure}

Let us consider weak, monochromatic, off-resonant, planar light incident upon a non-conducting fluid of small, diamagnetic, chiral molecules, as illustrated in Fig. \ref{Fig1}. Fluctuations of the optical properties within the scattering volume give rise to Rayleigh-Brillouin scattering away from the forward direction \cite{Rayleigh71a, Rayleigh71b, Rayleigh71c, Rayleigh81a, Rayleigh99a, Smoluchowski08a, Rayleigh10a, Einstein10a, Brillouin14a, Cabannes15a, Strutt18a, Rayleigh18a, Cabannes21a, Brillouin22a, Mandelstam26a, Cabannes28a, Cabannes28b, Raman28a, Cabannes29a, Manneback30a, Manneback30b, Gross30a, Landau34a}, a fraction of which is analysed at a detector in the far field. In what follows, we derive expressions for dimensionless circular spectral differentials and dimensionless circular intensity differentials which serve as convenient measures of the Rayleigh-Brillouin optical activity exhibited by the sample.

Our derivation borrows heavily from theoretical descriptions of Rayleigh-Brillouin scattering by Landau and Lifshitz \cite{Landau84a} and Berne and Pecora \cite{Berne00a} and is essentially an amalgamation of these with Barron's mechanism of Rayleigh optical activity \cite{Atkins69a, Barron71a, Barron04a}. See also \cite{Komarov63a, Young81a, Young82a, Fabelinskii94a}. We work in an inertial frame of reference with time $t$ and position vector $\mathbf{r}=x\hat{\mathbf{x}}+y\hat{\mathbf{y}}+z\hat{\mathbf{z}}$, where $x$, $y$ and $z$ are right-handed Cartesian coordinates and $\hat{\mathbf{x}}$, $\hat{\mathbf{y}}$ and $\hat{\mathbf{z}}$ are the associated unit vectors. The Einstein summation convention is to be understood with respect to unprimed Greek indices $\alpha,\beta,\dots\in\{x,y,z\}$ and primed Greek indices $\alpha^\prime,\beta^\prime,\dots\in\{X^{(n)},Y^{(n)},Z^{(n)}\}$, where $X^{(n)}$, $Y^{(n)}$ and $Z^{(n)}$ are molecule-fixed Cartesian coordinates for the $n^\mathrm{th}$ molecule. Complex quantities are decorated with tildes and unit vectors are decorated with carets. We use SI units throughout.

\subsection{The sample}
\label{The sample}
Within the scattering volume $V$, we model the sample as a collection of vibronically polarisable molecules that can translate with position vectors $\mathbf{R}^{(n)}=\mathbf{R}^{(n)}(t)$ and rotate with Euler angles $\vartheta^{(n)}=\vartheta^{(n)}(t)$, $\varphi^{(n)}=\varphi^{(n)}(t)$ and $\chi^{(n)}=\chi^{(n)}(t)$ ($n\in\{1,\dots\}$) \cite{Bunker05a}. In the interests of generality, we say nothing about the explicit forms of the positions and orientations of the molecules, except that they are such that the sample is optically homogeneous and isotropic on average. 

We take the light to satisfy Maxwell's equations in the form
\begin{align}
\boldsymbol{\nabla}\cdot\tilde{\mathbf{D}}=0, \ \ \ \ 
\boldsymbol{\nabla}\cdot\tilde{\mathbf{B}}=0, \ \ \ \ 
\boldsymbol{\nabla}\times\tilde{\mathbf{E}}=-\frac{\partial\tilde{\mathbf{B}}}{\partial t} \ \ \ \ \boldsymbol{\nabla}\times\tilde{\mathbf{B}}=\mu_0\frac{\partial\tilde{\mathbf{D}}}{\partial t} \label{Maxwells}
\end{align}
together with the constitutive relation
\begin{align}
\tilde{\mathbf{D}}&\approx\epsilon_0\tilde{\mathbf{E}}+\tilde{\mathbf{P}}-\frac{1}{\mathrm{i}\omega}\boldsymbol{\nabla}\times\tilde{\mathbf{M}}, \nonumber
\end{align}
where $\tilde{\mathbf{D}}=\tilde{\mathbf{D}}(\mathbf{r},t)$, $\tilde{\mathbf{E}}=\tilde{\mathbf{E}}(\mathbf{r},t)$, $\tilde{\mathbf{B}}=\tilde{\mathbf{B}}(\mathbf{r},t)$, $\tilde{\mathbf{P}}=\tilde{\mathbf{P}}(\mathbf{r},t)$ and $\tilde{\mathbf{M}}=\tilde{\mathbf{M}}(\mathbf{r},t)$ are the complex displacement, electric, magnetic, polarisation and magnetisation fields and $\omega$ is the angular frequency of the incident light \cite{Craig98a, Jackson98a, Barron04a}.

Working in the domain of linear optics to first order in multipolar expansions whilst neglecting local field corrections, we take
\begin{align}
\tilde{P}_\alpha\approx\sum_n \tilde{\mu}^{(n)}_\alpha\delta^3(\mathbf{r}-\mathbf{R}_n)-\sum_n\frac{1}{3}\tilde{\Theta}^{(n)}_{\alpha\beta}\partial_\beta\delta^3(\mathbf{r}-\mathbf{R}_n) \ \ \ \ \tilde{M}_\alpha\approx\sum_n\tilde{m}^{\prime(n)}_\alpha\delta^3(\mathbf{r}-\mathbf{R}_n) \nonumber
\end{align}
with 
\begin{align}
\tilde{\mu}^{(n)}_\alpha\approx\alpha^{(n)}_{\alpha\beta}\tilde{E}^{(n)}_\beta-\frac{1}{\omega}\epsilon_{\delta\gamma\beta}G^{\prime(n)}_{\alpha\delta}\partial_\gamma\tilde{E}^{(n)}_\beta+\frac{1}{3}A^{(n)}_{\alpha,\beta\gamma}\partial_\gamma\tilde{E}^{(n)}_\beta, \ \ \ \ \tilde{\Theta}^{(n)}_{\alpha\beta}\approx A^{(n)}_{\gamma,\alpha\beta}\tilde{E}^{(n)}_\gamma \ \ \ \ \tilde{m}^{\prime(n)}_\alpha\approx\mathrm{i}G^{\prime(n)}_{\beta\alpha}\tilde{E}^{(n)}_\beta, \nonumber 
\end{align}
where $\tilde{\mu}^{(n)}_\alpha=\tilde{\mu}^{(n)}_\alpha(t)$, $\tilde{\Theta}^{(n)}_{\alpha\beta}=\tilde{\Theta}^{(n)}_{\alpha\beta}(t)$ and $\tilde{m}^{\prime(n)}_\alpha=\tilde{m}^{\prime(n)}_\alpha(t)$ are the complex electric dipole, electric quadrupole and magnetic dipole moments of the $n^\mathrm{th}$ molecule and $\alpha^{(n)}_{\alpha\beta}=\alpha^{(n)}_{\beta\alpha}=\alpha^{(n)}_{\alpha\beta}(t)$, $G^{\prime(n)}_{\alpha\beta}=G^{\prime(n)}_{\alpha\beta}(t)$ and $A^{(n)}_{\alpha,\beta\gamma}=A^{(n)}_{\alpha,\gamma\beta}=A^{(n)}_{\alpha,\beta\gamma}(t)$ are the vibronic\footnote{By ``vibronic'', we mean simply that the polarisabilities account for electromagnetic perturbation of the vibrational and electronic degrees of freedom of the molecules. The initial and final states in the polarisabilities are taken to be the same; the molecules do \textit{not} undergo real vibronic transitions.} electric dipole-electric dipole, electric dipole-magnetic dipole and electric dipole-electric quadrupole polarisability tensors of the $n^\mathrm{th}$ molecule \cite{Craig98a, Barron04a}. The laboratory-fixed components of the polarisability tensors are related to the molecule-fixed components via relations like 
\begin{align}
\alpha_{\alpha\beta}^{(n)}&=\ell^{(n)}_{\alpha\alpha^\prime}\ell^{(n)}_{\beta\beta^\prime}\alpha_{\alpha^\prime\beta^\prime}^{(n)}, \nonumber
\end{align}
for example, where $\ell^{(n)}_{\alpha\alpha^\prime}=\ell^{(n)}_{\alpha\alpha^\prime}(\vartheta^{(n)}(t),\varphi^{(n)}(t),\chi^{(n)}(t))$ is the direction cosine tensor for the $n^\mathrm{th}$ molecule \cite{Barron04a, Bunker05a}.

\subsection{Incident light}
We take the incident light (superscript $\mathrm{i}$) to satisfy Maxwell's equations in the form
\begin{align}
\boldsymbol{\nabla}\cdot\tilde{\mathbf{D}}^\mathrm{i}=0, \ \ \ \ \boldsymbol{\nabla}\cdot\tilde{\mathbf{B}}^\mathrm{i}=0, \ \ \ \ \boldsymbol{\nabla}\times\tilde{\mathbf{E}}^\mathrm{i}=-\frac{\partial\tilde{\mathbf{B}}^\mathrm{i}}{\partial t} \ \ \ \ 
\boldsymbol{\nabla}\times\tilde{\mathbf{B}}^\mathrm{i}=\mu_0\frac{\partial\tilde{\mathbf{D}}^\mathrm{i}}{\partial t} \label{IncidentMaxwells}
\end{align}
together with the constitutive relation
\begin{align}
\tilde{\mathbf{D}}^\mathrm{i}&\approx\epsilon\tilde{\mathbf{E}}^\mathrm{i}-\gamma\boldsymbol{\nabla}\times\tilde{\mathbf{E}}^\mathrm{i}, \label{IncidentDisplacementConstitutive}
\end{align}
where $\tilde{\mathbf{D}}^\mathrm{i}=\tilde{\mathbf{D}}^\mathrm{i}(\mathbf{r},t)$, $\tilde{\mathbf{E}}^\mathrm{i}=\tilde{\mathbf{E}}^\mathrm{i}(\mathbf{r},t)$, $\tilde{\mathbf{B}}^\mathrm{i}=\tilde{\mathbf{B}}^\mathrm{i}(\mathbf{r},t)$ are the complex displacement, electric and magnetic fields of the incident light, $\epsilon$ is the average permittivity of the sample and $\gamma$ is the average optical activity parameter of the sample \cite{Craig98a, Jackson98a, Barron04a}. The solutions of (\ref{IncidentMaxwells}) and (\ref{IncidentDisplacementConstitutive}) are circularly polarised plane electromagnetic waves and superpositions thereof \cite{Barron04a}.

Let us assume that the incident light propagates only a short distance through the sample, as will typically be the case in an experiment using a small cuvette, for example. Accordingly, we neglect the optical rotation of the incident light ($\gamma\rightarrow 0$) and take
\begin{align}
\tilde{\mathbf{E}}^\mathrm{i}\approx E^{(0)}\tilde{\mathbf{e}}^\mathrm{i}\mathrm{e}^{\mathrm{i}(\boldsymbol{\kappa}^\mathrm{i}\cdot\mathbf{r}-\omega t)}, \ \ \ \ \boldsymbol{\kappa}^\mathrm{i}=\frac{\omega\mathbf{n}^\mathrm{i}}{c} \ \ \ \ \mathbf{n}^\mathrm{i}=n\hat{\mathbf{z}} \nonumber
\end{align}
with
\begin{align}
\hat{\mathbf{e}}^\mathrm{H}=\hat{\mathbf{y}}, \ \ \ \ \hat{\mathbf{e}}^\mathrm{V}=\hat{\mathbf{x}}, \ \ \ \ \tilde{\mathbf{e}}^\mathrm{L}=\frac{1}{\sqrt{2}}(\hat{\mathbf{x}}+\mathrm{i}\hat{\mathbf{y}}) \ \ \ \ \tilde{\mathbf{e}}^\mathrm{R}=\frac{1}{\sqrt{2}}(\hat{\mathbf{x}}-\mathrm{i}\hat{\mathbf{y}}), \nonumber 
\end{align}
where $E^{(0)}$, $\tilde{\mathbf{e}}^\mathrm{i}$, $\boldsymbol{\kappa}^\mathrm{i}$ and $\mathbf{n}^\mathrm{i}$ are the electric-field amplitude, complex polarisation vector, wavevector and propagation vector of the incident light and $n=\sqrt{\epsilon/\epsilon_0}$ is the average refractive index of the sample with $\gamma=0$.

\subsection{Scattered light}
We take the scattered light (superscript $\mathrm{s}$) to be the difference between the light and the incident light defined above, as
\begin{align}
\tilde{\mathbf{D}}^\mathrm{s}=\tilde{\mathbf{D}}-\tilde{\mathbf{D}}^\mathrm{i}, \ \ \ \
\tilde{\mathbf{E}}^\mathrm{s}=\tilde{\mathbf{E}}-\tilde{\mathbf{E}}^\mathrm{i} \ \ \ \ 
\tilde{\mathbf{B}}^\mathrm{s}=\tilde{\mathbf{B}}-\tilde{\mathbf{B}}^\mathrm{i}, \ \ \ \ 
\label{ScatteredFieldsDefinition}
\end{align}
where $\tilde{\mathbf{D}}^\mathrm{s}=\tilde{\mathbf{D}}^\mathrm{s}(\mathbf{r},t)$, $\tilde{\mathbf{E}}^\mathrm{s}=\tilde{\mathbf{E}}^\mathrm{s}(\mathbf{r},t)$, $\tilde{\mathbf{B}}^\mathrm{s}=\tilde{\mathbf{B}}^\mathrm{s}(\mathbf{r},t)$ are the complex displacement, electric and magnetic fields of the scattered light \cite{Landau84a, Berne00a}. 

Using (\ref{IncidentDisplacementConstitutive}) and (\ref{ScatteredFieldsDefinition}), we obtain
\begin{align}
\tilde{\mathbf{D}}^\mathrm{s}&\approx\epsilon\tilde{\mathbf{E}}^\mathrm{s}-\gamma\pmb{\nabla}\times\tilde{\mathbf{E}}^\mathrm{s}+\delta\tilde{\mathbf{D}} \label{ScatteredDisplacement}
\end{align}
with
\begin{align}
\delta\tilde{\mathbf{D}}&\approx\tilde{\mathbf{D}}-(\epsilon\tilde{\mathbf{E}}-\gamma\boldsymbol{\nabla}\times\tilde{\mathbf{E}}), \nonumber
\end{align}
where $\delta\tilde{\mathbf{D}}=\delta\tilde{\mathbf{D}}(\mathbf{r},t)$ embodies fluctuations with respect to the average optical properties of the sample, being the difference between the displacement field $\tilde{\mathbf{D}}$ and the form it would have if the sample had homogeneous and isotropic constitutive relations ($\epsilon\tilde{\mathbf{E}}-\gamma\boldsymbol{\nabla}\times\tilde{\mathbf{E}}$).

Using (\ref{Maxwells}), (\ref{IncidentMaxwells}), (\ref{ScatteredFieldsDefinition}) and (\ref{ScatteredDisplacement}) as well as the vector identity
\begin{align}
\boldsymbol{\nabla}\times(\boldsymbol{\nabla}\times\tilde{\mathbf{D}}^\mathrm{s})&=-\nabla^2\tilde{\mathbf{D}}^\mathrm{s}+\boldsymbol{\nabla}(\boldsymbol{\nabla}\cdot\tilde{\mathbf{D}}^\mathrm{s}), \nonumber
\end{align}
we find that $\tilde{\mathbf{D}}^\mathrm{s}$ must satisfy the wave equation
\begin{align}
\nabla^2\tilde{\mathbf{D}}^\mathrm{s}-\frac{n^2}{c^2}\frac{\partial^2\tilde{\mathbf{D}}^\mathrm{s}}{\partial t^2}+\mu_0\gamma\frac{\partial^2(\pmb{\nabla}\times\tilde{\mathbf{D}}^\mathrm{s})}{\partial t^2}&\approx-\pmb{\nabla}\times(\pmb{\nabla}\times\delta\tilde{\mathbf{D}}), \label{WaveEquation}
\end{align}
which shows that the fluctuations embodied by the difference $\delta\tilde{\mathbf{D}}$ drive waves in the displacement $\tilde{\mathbf{D}}^\mathrm{s}$ of the scattered light.

Neglecting multiple scattering\footnote{The condition $4\pi\mu_0^2\omega^4\alpha^2 NV^{1/3}\ll1$ should be well satisfied in the visible domain by typical small-molecule liquids for a cuvette of volume $V=1\,\mathrm{cm}^3$, say.}, we take 
\begin{align}
\delta\tilde{\mathbf{D}}\approx\delta\tilde{\mathbf{D}}^\mathrm{i} \label{ApproximatedeltaD1}
\end{align}
with
\begin{align}
\delta\tilde{D}^\mathrm{i}_\alpha&=\epsilon_0\tilde{E}^\mathrm{i}_\alpha+\sum_n\alpha^{(n)}_{\alpha\beta}\delta^3(\mathbf{r}-\mathbf{R}_n)\tilde{E}^{\mathrm{i}(n)}_\beta-\sum_n\left(\frac{1}{\omega}\epsilon_{\delta\gamma\beta}G^{\prime(n)}_{\alpha\delta}-\frac{1}{3}A^{(n)}_{\alpha,\beta\gamma}\right)\delta^3(\mathbf{r}-\mathbf{R}_n)\partial_\gamma\tilde{E}^{\mathrm{i}(n)}_\beta \nonumber \\
&+\sum_n\left(\frac{1}{\omega}\epsilon_{\delta\gamma\alpha}G^{\prime(n)}_{\beta\delta}-\frac{1}{3}A^{(n)}_{\beta,\alpha\gamma}\right)\partial_\gamma\delta^3(\mathbf{r}-\mathbf{R}_n)\tilde{E}^{\mathrm{i}(n)}_\beta-(\epsilon\tilde{E}^\mathrm{i}_\alpha+\gamma\epsilon_{\alpha\beta\gamma}\partial_\gamma\tilde{E}^\mathrm{i}_\beta), \label{deltaDiExplicit}
\end{align}
where $\delta\tilde{\mathbf{D}}^\mathrm{i}=\delta\tilde{\mathbf{D}}^\mathrm{i}(\mathbf{r},t)$ embodies fluctuations coupled directly to the incident light.

Let us assume that the scattered light (like the incident light) propagates only a short distance through the sample. Accordingly, we neglect the optical rotation of the scattered light ($\gamma=0$) and take the solution of (\ref{WaveEquation}) with (\ref{ApproximatedeltaD1}) to be
\begin{align}
\tilde{\mathbf{D}}^\mathrm{s}\approx\frac{1}{4\pi}\pmb{\nabla}\times\left\{\pmb{\nabla}\times\left[\iiint_V\frac{\delta\tilde{\mathbf{D}}^\mathrm{i}(\mathbf{r}^\prime,t^\prime)}{|\mathbf{r}-\mathbf{r}^\prime|}\mathrm{d}^3\mathbf{r}^\prime\right]\right\}, \label{DelayedSolution}
\end{align}
where $t^\prime=t-n|\mathbf{r}-\mathbf{r}^\prime|/c$ is the delayed time \cite{Jackson98a, Barron04a}. Let us assume moreover that the fluctuations are slow relative to the angular frequency $\omega$ of the incident light. Accordingly, we simplify (\ref{DelayedSolution}) by taking
\begin{align}
\tilde{\mathbf{D}}^\mathrm{s}&\approx\frac{1}{4\pi}\pmb{\nabla}\times\left\{\pmb{\nabla}\times\left[\iiint_V\frac{\delta\tilde{\mathbf{D}}^\mathrm{i}(\mathbf{r}^\prime,t)\mathrm{e}^{\mathrm{i}\omega t}\mathrm{e}^{-\mathrm{i}\omega t^\prime}}{|\mathbf{r}-\mathbf{r}^\prime|}\mathrm{d}^3\mathbf{r}^\prime\right]\right\}, \label{PseudoDelayedSolution}
\end{align}
where we have effectively separated the fast and slow time dependencies of $\delta\tilde{\mathbf{D}}^\mathrm{i}$ by writing 
\begin{align}
\delta\tilde{\mathbf{D}}^\mathrm{i}(\mathbf{r}^\prime,t^\prime)=[\delta\tilde{\mathbf{D}}^\mathrm{i}(\mathbf{r}^\prime,t^\prime)\exp(\mathrm{i}\omega t^\prime)]\exp(-\mathrm{i}\omega t^\prime)\rightarrow[\delta\tilde{\mathbf{D}}^\mathrm{i}(\mathbf{r}^\prime,t)\exp(\mathrm{i}\omega t)]\exp(-\mathrm{i}\omega t^\prime), \nonumber
\end{align}
thus retaining delay for the fast dependencies only \cite{Landau84a, Berne00a}.

\subsection{Analysed signal at the detector}
Let us focus now on the form of the scattered light at the detector (superscript $\mathrm{d}$). We take the $y$-$z$ plane to be the scattering plane, without loss of generality. It is convenient to introduce unit vectors $\hat{\mathbf{x}}^\mathrm{d}=\hat{\mathbf{x}}$, $\hat{\mathbf{y}}^\mathrm{d}=\cos\theta\hat{\mathbf{y}}-\sin\theta\hat{\mathbf{z}}$ and $\hat{\mathbf{z}}^\mathrm{d}=\sin\theta\hat{\mathbf{y}}+\cos\theta\hat{\mathbf{z}}$ aligned with the detector, where $\theta$ is the scattering angle. 

As the detector lies in the far field, we take
\begin{align}
\frac{1}{|\mathbf{r}-\mathbf{r}^\prime|}\approx\frac{1}{R} \ \ \ \omega t^\prime\approx \omega t-\kappa R+\boldsymbol{\kappa}^\mathrm{d}\cdot\mathbf{r}^\prime \label{FarField}
\end{align}
with
\begin{align}
\kappa=\frac{\omega n}{c}, \ \ \ \ \boldsymbol{\kappa}^\mathrm{d}=\frac{\omega\mathbf{n}^\mathrm{d}}{c} \ \ \ \ \mathbf{n}^\mathrm{d}=n\hat{\mathbf{z}}^\mathrm{d}, \label{ScatteredWavevector}
\end{align}
where $R$ is the distance from the centre of the scattering volume $V$ to the detector ($R\gg V^{1/3}$, $\kappa R\gg 1$) and $\kappa$, $\boldsymbol{\kappa}^\mathrm{d}$ and $\mathbf{n}^\mathrm{d}$ are the angular wavenumber, wavevector and propagation vector of the scattered light at the detector \cite{Jackson98a, Barron04a}. Using (\ref{PseudoDelayedSolution}) and (\ref{FarField}) as well as the vector identity
\begin{align}
-\boldsymbol{\kappa}^\mathrm{d}\times(\boldsymbol{\kappa}^\mathrm{d}\times\mathbf{V})&=(\boldsymbol{\kappa}^\mathrm{d}\cdot\boldsymbol{\kappa}^\mathrm{d})\mathbf{V}-\boldsymbol{\kappa}^\mathrm{d}(\boldsymbol{\kappa}^\mathrm{d}\cdot\mathbf{V}) \nonumber
\end{align}
whilst retaining only the leading contributions which fall off as $1/R$, we find that the complex electric field $\tilde{\mathbf{E}}^{\mathrm{d}\mathrm{i}}=\tilde{\mathbf{E}}^{\mathrm{d}\mathrm{i}}(t)$ of the scattered light at the detector has the form
\begin{align}
\tilde{E}^{\mathrm{d}\mathrm{i}}_\alpha&\approx\frac{\mu_0\omega^2}{4\pi R}\mathrm{e}^{\mathrm{i}(\kappa R-\omega t)}(\delta_{\alpha\beta}-\hat{n}^\mathrm{d}_\alpha\hat{n}^\mathrm{d}_\beta)\iiint_V\delta\tilde{D}^\mathrm{i}_\beta(\mathbf{r}^\prime,t)\mathrm{e}^{\mathrm{i}\omega t}\mathrm{e}^{-\mathrm{i}\boldsymbol{\kappa}^\mathrm{d}\cdot\mathbf{r}^\prime}\mathrm{d}^3\mathbf{r}^\prime, \label{ScatteredDNotIntegrated}
\end{align}
where we have taken $\tilde{\mathbf{E}}^{\mathrm{d}\mathrm{i}}\approx\tilde{\mathbf{D}}^{\mathrm{d}\mathrm{i}}/\epsilon$, $\tilde{\mathbf{D}}^{\mathrm{d}\mathrm{i}}=\tilde{\mathbf{D}}^{\mathrm{d}\mathrm{i}}(t)$ being the complex displacement of the scattered light at the detector. Substituting (\ref{deltaDiExplicit}) explicitly into (\ref{ScatteredDNotIntegrated}) then integrating by parts and neglecting boundary terms as well as forward-scattering contributions, we obtain
\begin{align}
\tilde{E}^{\mathrm{d}\mathrm{i}}_{\alpha}&\approx\frac{\mu_0\omega^2E^{(0)}}{4\pi R}\mathrm{e}^{\mathrm{i}(\kappa R-\omega t)}(\delta_{\alpha\beta}-\hat{n}^\mathrm{d}_\alpha\hat{n}^\mathrm{d}_\beta)\Bigg[
\tilde{\alpha}_{\beta\gamma}\tilde{e}^\mathrm{i}_\gamma-\frac{\mathrm{i}}{c}(\tilde{G}_{\beta\epsilon}^\prime\epsilon_{\epsilon\delta\gamma}\tilde{e}^\mathrm{i}_\gamma n^\mathrm{i}_\delta-\tilde{G}_{\gamma\epsilon}^\prime\epsilon_{\epsilon\delta\beta}\tilde{e}^\mathrm{i}_\gamma n^\mathrm{d}_\delta) \nonumber \\
&+\frac{\mathrm{i}\omega}{3c}(\tilde{A}_{\beta,\gamma\delta}\tilde{e}^\mathrm{i}_\gamma n^\mathrm{i}_\delta-\tilde{A}_{\gamma,\beta\delta}\tilde{e}^\mathrm{i}_\gamma n^\mathrm{d}_\delta)\Bigg] \label{DetectedElectricField}
\end{align}
with
\begin{align}
\tilde{\alpha}_{\alpha\beta}=\sum_n\alpha_{\alpha\beta}^{(n)}\mathrm{e}^{-\mathrm{i}\mathbf{q}\cdot\mathbf{R}_n}, \ \ \ \ \tilde{G}^\prime_{\alpha\beta}=\sum_nG^{\prime(n)}_{\alpha\beta}\mathrm{e}^{-\mathrm{i}\mathbf{q}\cdot\mathbf{R}_n} \ \ \ \ \tilde{A}_{\alpha,\beta\gamma}=\sum_nA^{(n)}_{\alpha,\beta\gamma}\mathrm{e}^{-\mathrm{i}\mathbf{q}\cdot\mathbf{R}_n}, \nonumber
\end{align}
where $\tilde{\alpha}_{\alpha\beta}=\tilde{\alpha}_{\alpha\beta}(t)$, $\tilde{G}^\prime_{\alpha\beta}=\tilde{G}^\prime_{\alpha\beta}(t)$ and $\tilde{A}_{\alpha,\beta\gamma}=\tilde{A}_{\alpha,\beta\gamma}(t)$ are spatial Fourier transforms of polarisability densities, evaluated at the wavevector difference $\mathbf{q}=\boldsymbol{\kappa}^\mathrm{d}-\boldsymbol{\kappa}^\mathrm{i}\ne0$ \cite{Landau84a, Berne00a}. Note that $q=|\mathbf{q}|=2\omega n\sin(\theta/2)/c$.

We take the analysed signal $\tilde{A}^{\mathrm{d}\mathrm{i}}=\tilde{A}^{\mathrm{d}\mathrm{i}}(t)$ at the detector to be
\begin{align}
\tilde{A}^{\mathrm{d}\mathrm{i}}&=\tilde{a}_\alpha^{\mathrm{d}\ast}\tilde{E}_\alpha^{\mathrm{d}\mathrm{i}} \label{AnalysedSignal} 
\end{align} 
with
\begin{align}
\hat{\mathbf{a}}^\mathrm{H}=\hat{\mathbf{y}}^\mathrm{s}, \ \ \ \ \hat{\mathbf{a}}^\mathrm{V}=\hat{\mathbf{x}}^\mathrm{s}, \ \ \ \ 
\tilde{\mathbf{a}}^\mathrm{L}=\frac{1}{\sqrt{2}}(\hat{\mathbf{x}}^\mathrm{s}+\mathrm{i}\hat{\mathbf{y}}^\mathrm{s}) \ \ \ \
\tilde{\mathbf{a}}^\mathrm{R}=\frac{1}{\sqrt{2}}(\hat{\mathbf{x}}^\mathrm{s}-\mathrm{i}\hat{\mathbf{y}}^\mathrm{s}), \nonumber 
\end{align}
where $\tilde{\mathbf{a}}^\mathrm{d}$ is an analysation vector that picks off the desired polarisation component of the electric field $\tilde{\mathbf{E}}^{\mathrm{d}\mathrm{i}}$.

\subsection{Frequency spectrum and total intensity}
\label{Frequency spectrum and total intensity}
The frequency spectrum $\mathbb{I}^{\mathrm{d}\mathrm{i}}=\mathbb{I}^{\mathrm{d}\mathrm{i}}(\Omega)$ of the analysed signal $\tilde{A}^{\mathrm{d}\mathrm{i}}$ can be calculated as the temporal Fourier transform of the autocorrelation of $\tilde{A}^{\mathrm{d}\mathrm{i}}$, as 
\begin{align}
\mathbb{I}^{\mathrm{d}\mathrm{i}}&=\frac{1}{2\pi}\int_{-\infty}^\infty\langle\tilde{A}^{\mathrm{d}\mathrm{i}\ast}(t)\tilde{A}^{\mathrm{d}\mathrm{i}}(t+\tau)\rangle\mathrm{e}^{\mathrm{i}\Omega\tau}\mathrm{d}\tau, \label{SpectralDensity}
\end{align}
where the angular brackets denote a time average, $\Omega$ is an angular frequency and $\tau$ is a correlation time \cite{Berne00a}. The total intensity $I^{\mathrm{d}\mathrm{i}}$ of the analysed signal can then be calculated as the integral of the frequency spectrum $\mathbb{I}^{\mathrm{d}\mathrm{i}}$, as 
\begin{align}
I^{\mathrm{d}\mathrm{i}}&=\int_{-\infty}^\infty\mathbb{I}^{\mathrm{d}\mathrm{i}}(\Omega)\mathrm{d}\Omega. \nonumber 
\end{align}
Note that $\mathbb{I}^{\mathrm{d}\mathrm{i}}=0$ for $\Omega<0$, assuming that $\tilde{A}^{\mathrm{d}\mathrm{i}}=0$ for $\Omega<0$. 

\begin{table*}[!t]
\centering
\begin{tabular}{c|c|c|c|c|c|c|c|c|c}
$\mathrm{a}$ & $\mathrm{b}$ & $\mathrm{c}$ & $\mathrm{d}$ & $3c\mathbb{A}/nK$ & $3c\mathbb{B}/nK$ & $3c\mathbb{C}/nK$ & $3\mathbb{D}/K$ & $3\mathbb{E}/K$ & $3\mathbb{F}/K$  \\ \hline
R & H & L & H & $24\bbbeta^2_G-8\bbbeta^2_A$ & $180\bbalpha\mathbb{G}^\prime-20\bbbeta^2_G-12\bbbeta^2_A$ & $180\bbalpha\mathbb{G}^\prime+4\bbbeta^2_G+12\bbbeta^2_A$ & $12\bbbeta^2$ & $0$ & $90\bbalpha^2+2\bbbeta^2$ \\ \hline
R & V & L & V & $180\bbalpha\mathbb{G}^\prime+28\bbbeta^2_G+4\bbbeta^2_A$ & $180\bbalpha\mathbb{G}^\prime-20\bbbeta^2_G-12\bbbeta^2_A$ & $0$ & $90\bbalpha^2+14\bbbeta^2$ & $0$ & $0$ \\ \hline
R & N & L & N & $90\bbalpha\mathbb{G}^\prime+26\bbbeta^2_G-2\bbbeta^2_A$ & $180\bbalpha\mathbb{G}^\prime-20\bbbeta^2_G-12\bbbeta^2_A$ & $90\bbalpha\mathbb{G}^\prime+2\bbbeta^2_G+6\bbbeta^2_A$ & $45\bbalpha^2+13\bbbeta^2$ & $0$ & $45\bbalpha^2+\bbbeta^2$ \\ \hline
H & R & H & L & $24\bbbeta^2_G-8\bbbeta^2_A$ & $180\bbalpha\mathbb{G}^\prime-20\bbbeta^2_G-12\bbbeta^2_A$ & $180\bbalpha\mathbb{G}^\prime+4\bbbeta^2_G+12\bbbeta^2_A$ & $12\bbbeta^2$ & $0$ & $90\bbalpha^2+2\bbbeta^2$ \\ \hline
V & R & V & L & $180\bbalpha\mathbb{G}^\prime+28\bbbeta^2_G+4\bbbeta^2_A$ & $180\bbalpha\mathbb{G}^\prime-20\bbbeta^2_G-12\bbbeta^2_A$ & $0$ & $90\bbalpha^2+14\bbbeta^2$ & $0$ & $0$ \\ \hline
N & R & N & R & $90\bbalpha\mathbb{G}^\prime+26\bbbeta^2_G-2\bbbeta^2_A$ & $180\bbalpha\mathbb{G}^\prime-20\bbbeta^2_G-12\bbbeta^2_A$ & $90\bbalpha\mathbb{G}^\prime+2\bbbeta^2_G+6\bbbeta^2_A$ & $45\bbalpha^2+13\bbbeta^2$ & $0$ & $45\bbalpha^2+\bbbeta^2$  \\ \hline
R & R & L & L & $180\bbalpha \mathbb{G}^\prime+52\bbbeta_G^2-4\bbbeta_A^2$ & $360\bbalpha\mathbb{G}^\prime-40\bbbeta_G^2-24\bbbeta_A^2$ & $180\bbalpha\mathbb{G}^\prime+4\bbbeta_G^2+12\bbbeta_A^2$ & $45\bbalpha^2+13\bbbeta^2$ & $90\bbalpha^2-10\bbbeta^2$ & $45\bbalpha^2+\bbbeta^2$ 
\end{tabular}
\caption{Coefficients for some important SCP, ICP and DCP$_\mathrm{I}$ configurations. The $\mathrm{N}$ results are averages of the relevant $\mathrm{H}$ and $\mathrm{V}$ results, the $\mathrm{N}$ standing for ``natural'' (unpolarised incident light or unanalysed scattered light). Note that the entries in rows one through three match the entries in rows four through six, in accord with the principle of reciprocity \cite{Rayleigh00a, Krishnan38a, Perrin42a, Barron04a}.}
\label{Table1}
\end{table*}

Substituting (\ref{DetectedElectricField}) and (\ref{AnalysedSignal}) into (\ref{SpectralDensity}) whilst working to first order in multipolar expansions, we obtain
\begin{align}
\mathbb{I}^{\mathrm{d}\mathrm{i}}&\approx\frac{30 K}{2\pi}\int_{-\infty}^\infty\langle\tilde{\alpha}^{\mathrm{d}\mathrm{i}\ast}(t)\tilde{\alpha}^{\mathrm{d}\mathrm{i}}(t+\tau)+\tilde{\alpha}^{\mathrm{d}\mathrm{i}\ast}(t)\tilde{\chi}^{\mathrm{d}\mathrm{i}}(t+\tau)+\tilde{\alpha}^{\mathrm{d}\mathrm{i}}(t+\tau)\tilde{\chi}^{\mathrm{d}\mathrm{i}\ast}(t)\rangle\mathrm{e}^{\mathrm{i}(\Omega-\omega)\tau}\mathrm{d}\tau \label{SpectralDensity2}
\end{align} 
with
\begin{align}
K&=\frac{1}{30}\left(\frac{\mu_0\omega^2E^{(0)}}{4\pi R}\right)^2, \nonumber \\
\tilde{\alpha}^{\mathrm{d}\mathrm{i}}&=\tilde{a}^{\mathrm{d}\ast}_\alpha\tilde{\alpha}_{\alpha\beta}\tilde{e}^\mathrm{i}_\beta \nonumber \\
\tilde{\chi}^{\mathrm{d}\mathrm{i}}&=-\frac{\mathrm{i}}{c}(\tilde{a}^{\mathrm{d}\ast}_\alpha \tilde{G}^\prime_{\alpha\delta}\epsilon_{\delta\gamma\beta}\tilde{e}^\mathrm{i}_\beta n^\mathrm{i}_\gamma-\tilde{a}^{\mathrm{d}\ast}_\alpha\tilde{G}^\prime_{\beta\delta}\epsilon_{\delta\gamma\alpha}\tilde{e}^\mathrm{i}_\beta n^\mathrm{s}_\gamma)+\frac{\mathrm{i}\omega}{3c}(\tilde{a}^{\mathrm{d}\ast}_\alpha\tilde{A}_{\alpha,\beta\gamma}\tilde{e}^\mathrm{i}_\beta n^\mathrm{i}_\gamma-\tilde{a}^{\mathrm{d}\ast}_\alpha\tilde{A}_{\beta,\alpha\gamma}\tilde{e}^\mathrm{i}_\beta n^\mathrm{s}_\gamma), \nonumber
\end{align}
where $K$ is a prefactor that contains the usual $\omega^4$, $E^{(0)2}$ and $1/R^2$ scalings characteristic of Rayleigh scattering in the far field and $\tilde{\alpha}^{\mathrm{d}\mathrm{i}}=\tilde{\alpha}^{\mathrm{d}\mathrm{i}}(t)$ and $\tilde{\chi}^{\mathrm{d}\mathrm{i}}=\tilde{\chi}^{\mathrm{d}\mathrm{i}}(t)$ are convenient shorthands.

\subsection{Dimensionless circular spectral and intensity differentials}
\label{Dimensionless circular spectral density and intensity differentials}
RayOA (and by extension RayBOA) can manifest as an intensity difference with respect to left- and right-handed circular polarisation states in the scattered light (scattered circular polarisation or SCP) \cite{Atkins69a}, the incident light (incident circular polarisation or ICP) \cite{Barron71a} or both simultaneously (dual circular polarisation or DCP$_\mathrm{I}$) \cite{Nafie89a}. 

As convenient measures of RayBOA, we identify dimensionless circular spectral differentials $\bbDelta=\bbDelta(\Omega)$ of the form
\begin{align}
\bbDelta&=\frac{\mathbb{I}^{\mathrm{a}\mathrm{b}}-\mathbb{I}^{\mathrm{c}\mathrm{d}}}{\mathbb{I}^{\mathrm{a}\mathrm{b}}+\mathbb{I}^{\mathrm{c}\mathrm{d}}} \nonumber \\
&\approx\frac{\mathbb{A}+\mathbb{B}\cos\theta+\mathbb{C}\cos^2\theta}{\mathbb{D}+\mathbb{E}\cos\theta+\mathbb{F}\cos^2\theta} \label{CircularSpectralDensityDifference}
\end{align}
and circular intensity differentials $\Delta$ of the form
\begin{align}
\Delta&=\frac{{I}^{\mathrm{a}\mathrm{b}}-{I}^{\mathrm{c}\mathrm{d}}}{{I}^{\mathrm{a}\mathrm{b}}+{I}^{\mathrm{c}\mathrm{d}}} \nonumber \\
&\approx\frac{\int_{-\infty}^\infty\mathbb{A}(\Omega)\mathrm{d}\Omega+\int_{-\infty}^\infty\mathbb{B}(\Omega)\mathrm{d}\Omega\cos\theta+\int_{-\infty}^\infty\mathbb{C}(\Omega)\mathrm{d}\Omega\cos^2\theta}{\int_{-\infty}^\infty\mathbb{D}(\Omega)\mathrm{d}\Omega+\int_{-\infty}^\infty\mathbb{E}(\Omega)\mathrm{d}\Omega\cos\theta+\int_{-\infty}^\infty\mathbb{F}(\Omega)\mathrm{d}\Omega\cos^2\theta}, \label{CircularIntensityDifference}
\end{align}
where $\mathrm{a}$ and $\mathrm{c}$ refer to analysed polarisation states of the scattered light, $\mathrm{b}$ and $\mathrm{d}$ refer to polarisation states of the incident light and $\mathbb{A}=\mathbb{A}(\Omega)$, $\mathbb{B}=\mathbb{B}(\Omega)$, $\mathbb{C}=\mathbb{C}(\Omega)$, $\mathbb{D}=\mathbb{D}(\Omega)$, $\mathbb{E}=\mathbb{E}(\Omega)$ and $\mathbb{F}=\mathbb{F}(\Omega)$ are coefficients that depend on the specific configuration being considered. Substituting (\ref{SpectralDensity2}) into (\ref{CircularSpectralDensityDifference}) and (\ref{CircularIntensityDifference}) and making use of basic symmetry arguments, we obtain the results listed in Table \ref{Table1} with
\begin{align}
\bbalpha^2&=\frac{1}{2\pi}\int_{-\infty}^\infty\langle\frac{1}{9}\tilde{\alpha}^\ast_{\alpha\alpha}(t)\tilde{\alpha}_{\beta\beta}(t+\tau)\rangle\mathrm{e}^{\mathrm{i}(\Omega-\omega)\tau}\mathrm{d}\tau, \nonumber \\
\bbbeta^2&=\frac{1}{2\pi}\int_{-\infty}^\infty\langle\frac{1}{2}[3\tilde{\alpha}^\ast_{\alpha\beta}(t)\tilde{\alpha}_{\alpha\beta}(t+\tau)-\tilde{\alpha}^\ast_{\alpha\alpha}(t)\tilde{\alpha}_{\beta\beta}(t+\tau)]\rangle\mathrm{e}^{\mathrm{i}(\Omega-\omega)\tau}\mathrm{d}\tau, \nonumber \\
\bbalpha\mathbb{G}^\prime&=\frac{1}{2\pi}\int_{-\infty}^\infty\langle\frac{1}{18}[\tilde{\alpha}^\ast_{\alpha\alpha}(t)\tilde{G}^\prime_{\beta\beta}(t+\tau)+\tilde{\alpha}_{\alpha\alpha}(t+\tau)\tilde{G}^{\prime\ast}_{\beta\beta}(t)]\rangle\mathrm{e}^{\mathrm{i}(\Omega-\omega)\tau}\mathrm{d}\tau, \nonumber \\
\bbbeta^2_G&=\frac{1}{2\pi}\int_{-\infty}^\infty\langle\frac{1}{4}[3\tilde{\alpha}^\ast_{\alpha\beta}(t)\tilde{G}^\prime_{\alpha\beta}(t+\tau)-\tilde{\alpha}^\ast_{\alpha\alpha}(t)\tilde{G}^\prime_{\beta\beta}(t+\tau)+3\tilde{\alpha}_{\alpha\beta}(t+\tau)\tilde{G}^{\prime\ast}_{\alpha\beta}(t) -\tilde{\alpha}_{\alpha\alpha}(t+\tau)\tilde{G}^{\prime\ast}_{\beta\beta}(t)]\rangle\mathrm{e}^{\mathrm{i}(\Omega-\omega)\tau}\mathrm{d}\tau \nonumber \\
\bbbeta^2_A&=\frac{1}{2\pi}\int_{-\infty}^\infty\langle\frac{\omega}{4}[\epsilon_{\alpha\beta \gamma}\tilde{\alpha}^\ast_{\alpha\delta}(t)\tilde{A}_{\beta,\gamma\delta}(t+\tau)+\epsilon_{\alpha\beta\gamma}\tilde{\alpha}_{\alpha\delta}(t+\tau)\tilde{A}^\ast_{\beta,\gamma\delta}(t)]\rangle\mathrm{e}^{\mathrm{i}(\Omega-\omega)\tau}\mathrm{d}\tau, \label{Blackboards}
\end{align}
where $\bbalpha^2=\bbalpha^2(\Omega)$ accounts for isotropic electric dipole-electric dipole scattering, $\bbbeta^2=\bbbeta^2(\Omega)$ accounts for anisotropic electric dipole-electric dipole scattering, $\bbalpha\mathbb{G}^\prime=\bbalpha\mathbb{G}^\prime(\Omega)$ accounts for isotropic electric dipole-magnetic dipole scattering, $\bbbeta^2_G=\bbbeta^2_G(\Omega)$ accounts for anisotropic electric dipole-magnetic dipole scattering and $\bbbeta^2_A=\bbbeta^2_A(\Omega)$ accounts for anisotropic electric dipole-electric quadrupole scattering. Note that $\bbalpha^2$ and $\bbbeta^2$ are chirally insensitive whereas $\bbalpha\mathbb{G}^\prime$, $\bbbeta^2_G$ and $\bbbeta^2_A$ have equal magnitudes but opposite signs for enantiomorphic samples. It follows that both $\bbDelta$ and $\Delta$ have equal magnitudes but opposite signs for enantiomorphous samples, thus serving as signatures of chirality.
% We have assumed for now that these are real.

Let us emphasise here that the circular differentials $\bbDelta$ and $\Delta$ are not simply related to each other, in particular that
\begin{align}
\Delta\ne \int_{-\infty}^\infty \bbDelta(\Omega)\mathrm{d}\Omega. \nonumber
\end{align}
They provide different insights, as we will see below. 

As a quick check on the validity of our results, we note that for the special case of a single molecule held fixed at the origin, we have $\bbalpha^2\rightarrow\alpha^2\delta(\Omega-\omega)$, $\bbbeta^2\rightarrow\beta^2\delta(\Omega-\omega)$, 
$\bbalpha\mathbb{G}^\prime\rightarrow\alpha G^\prime\delta(\Omega-\omega)$, $\bbbeta_G^2\rightarrow\beta_G^2\delta(\Omega-\omega)$ and $\bbbeta_A^2\rightarrow\beta_A^2\delta(\Omega-\omega)$, where $\alpha^2$, $\beta^2$, $\alpha G^\prime$, $\beta_G^2$ and $\beta_A^2$ are the usual single-molecule invariants \cite{Barron04a, Zuber08a}. This sees our results for the circular intensity differentials $\Delta$ reduce immediately to the (rotationally averaged) results reported previously elsewhere \cite{Barron71a, Andrews80a, Barron04a, Zuber08a}, as they should.

%%%%%%%%%%%%%%%%%%%%%%%%%%%%%%%%%%%%%%
%      
%%%%%%%%%%%%%%%%%%%%%%%%%%%%%%%%%%%%%%
%%%%%%%%%%%%%%%%%%%%%%%%%%%%%%%%%%%%%%
%       TOY MODEL
%%%%%%%%%%%%%%%%%%%%%%%%%%%%%%%%%%%%%%
%%%%%%%%%%%%%%%%%%%%%%%%%%%%%%%%%%%%%%
%       
%%%%%%%%%%%%%%%%%%%%%%%%%%%%%%%%%%%%%%

\section{Toy model}
\label{Toy model}
For the sake of illustration, let us now evaluate the circular differentials $\bbDelta$ and $\Delta$ for a toy model of an enantiopure neat liquid. This model is not meant to provide accurate predictions for real liquids. Rather, we include it to demonstrate the mathematical extraction of spectra, the explicit application of our theory to real liquids being a challenging task that we will return to in future publications.

Considering molecules with quasi-cylindrical symmetry for the sake of simplicity, we take 
\begin{align}
\alpha^{(n)}_{\alpha\beta}&=\alpha\delta_{\alpha\beta}+\Delta\alpha\left(\hat{u}^{(n)}_\alpha\hat{u}^{(n)}_\beta-\frac{1}{3}\delta_{\alpha\beta}\right), \nonumber \\
G^{\prime(n)}_{\alpha\beta}&=G^\prime\delta_{\alpha\beta}+\Delta G^\prime\left(\hat{u}^{(n)}_\alpha\hat{u}^{(n)}_\beta-\frac{1}{3}\delta_{\alpha\beta}\right) \nonumber \\
A^{(n)}_{\alpha\beta}&=\frac{\omega}{2}\epsilon_{\alpha\gamma\delta}A^{(n)}_{\gamma,\beta\delta} \nonumber \\
&=\Delta A\left(\hat{u}^{(n)}_\alpha\hat{u}^{(n)}_\beta-\frac{1}{3}\delta_{\alpha\beta}\right), \label{Polarisabilities} 
\end{align}
where $\alpha$ and $G^\prime$ are isotropic polarisabilities; $\Delta \alpha$, $\Delta G^\prime$ and $\Delta A$ are polarisability anisotropies and $\hat{\mathbf{u}}^{(n)}=\hat{\mathbf{u}}^{(n)}(t)$ is a unit vector dictated by the orientation of the $n^\mathrm{th}$ molecule \cite{Berne00a, Barron04a}. Note that $\Delta\alpha=\Delta G^\prime=\Delta A=0$ for molecules with spherical rather than quasi-cylindrical symmetry. Substituting (\ref{Polarisabilities}) into (\ref{Blackboards}), we obtain
\begin{align}
\bbalpha^2=\alpha^2 S, \ \ \ \ \bbbeta^2=\beta^2\Theta, \ \ \ \ \bbalpha\mathbb{G}^\prime=\alpha G^\prime S, \ \ \ \ \bbbeta^2_G=\beta^2_G\Theta \ \ \ \ \bbbeta^2_A=\beta^2_A\Theta \nonumber
\end{align}
with
\begin{align}
S&=\frac{1}{2\pi}\int_{-\infty}^\infty \langle\sum_n\sum_m\mathrm{e}^{\mathrm{i}\mathbf{q}\cdot[\mathbf{R}_n(t)-\mathbf{R}_m(t+\tau)]} \rangle\mathrm{e}^{\mathrm{i}(\Omega-\omega)\tau}\mathrm{d}\tau \nonumber \\
\Theta&=\frac{1}{2\pi}\int_{-\infty}^\infty \langle\sum_n\sum_m\frac{1}{2}\{3[\hat{\mathbf{u}}^{(n)}(t)\cdot\hat{\mathbf{u}}^{(m)}(t+\tau)]^2-1\}\mathrm{e}^{\mathrm{i}\mathbf{q}\cdot[\mathbf{R}_n(t)-\mathbf{R}_m(t+\tau)]} \rangle\mathrm{e}^{\mathrm{i}(\Omega-\omega)\tau}\mathrm{d}\tau, \nonumber
\end{align}
where $\alpha^2$, $\beta^2$, $\alpha G^\prime$, $\beta^2=\Delta\alpha^2$, $\beta^2_G=\Delta\alpha\Delta G^\prime$ and $\beta^2_A=2\Delta\alpha\Delta A/3$ are the usual single-molecule invariants \cite{Barron04a, Zuber08a} and $S=S(\Omega)$ and $\Theta=\Theta(\Omega)$ are dynamic structure factors \cite{vanHove54a, Berne00a}. 

A simple hydrodynamic model (neglecting intramolecular relaxation) gives 
\begin{align}
S&\approx N^2 V\chi_Tk_\mathrm{B}T\left\{\left(1-\frac{1}{\gamma}\right)\frac{1}{\pi}\frac{D_T q^2}{(\Omega-\omega)^2+(D_T q^2)^2}+\frac{1}{2\gamma}\left[\frac{1}{\pi}\frac{\Gamma q^2}{(\Omega-\omega+vq)^2+(\Gamma q^2)^2}+\frac{1}{\pi}\frac{\Gamma q^2}{(\Omega-\omega-vq)^2+(\Gamma q^2)^2}\right]\right\}, \nonumber
\end{align}
where $N$ is the average number density, $\chi_T$ is the isothermal compressibility, $T$ is the temperature, $\gamma$ is the heat capacity ratio, $D_T$ is the thermal diffusivity, $v$ is the adiabatic speed of sound and $\Gamma$ is the classical sound attenuation coefficient \cite{Landau34a, Landau84a, Berne00a, Boyd03a}. A simple rotational diffusion model (neglecting translational effects and orientational correlations between molecules) gives 
\begin{align}
\Theta&\approx NV \frac{1}{\pi}\frac{\Gamma_\Theta}{(\Omega-\omega)^2+\Gamma_\Theta^2}, \nonumber
\end{align}
where $\Gamma_\Theta$ is a tumbling rate \cite{Debye29a, Berne00a}. The time-domain molecular dynamics that underpin these forms for $S$ and $\Theta$ are described explicitly in \cite{Debye29a, Landau34a, Landau84a, Berne00a, Boyd03a}.

\begin{figure}[h!]
\centering
\includegraphics[width=0.5\linewidth]{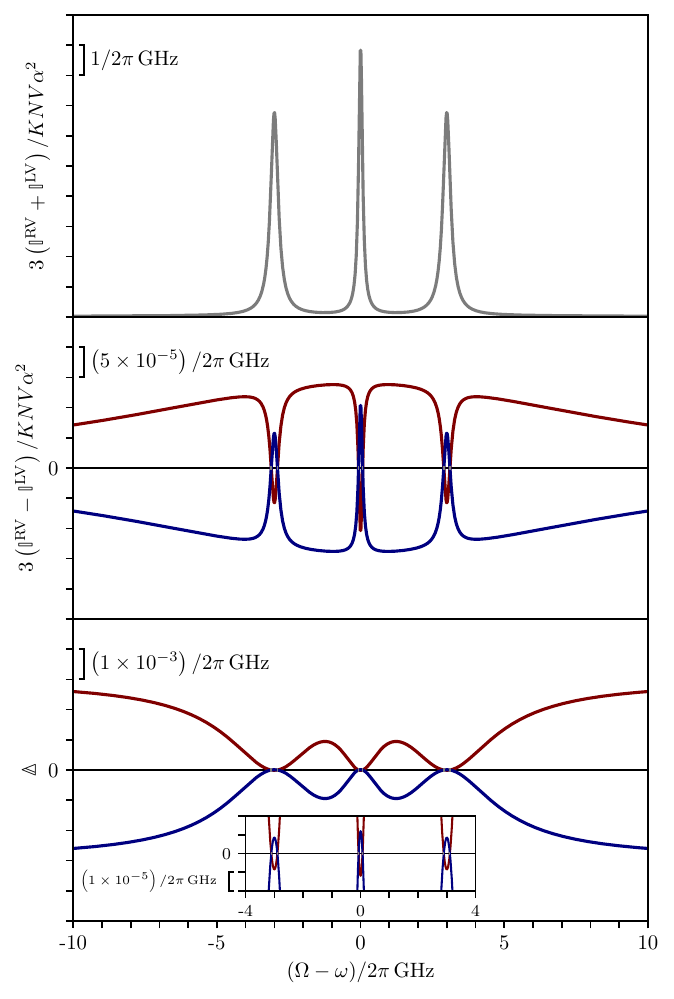}
\caption{\small Circular spectra predicted for a toy model of an enantiopure neat liquid. The red and blue curves correspond to opposite enantiomers.}
\label{Fig2}
\end{figure}

Focussing on a right-angled SCP configuration with vertically polarised incident light for the sake of concreteness, we have
\begin{align}
\bbDelta=\frac{\frac{nK}{3c}(180\alpha G^\prime S+28\beta^2_G\Theta+4\beta^2_A\Theta)}{\frac{K}{3}(90\alpha^2 S+14\beta^2\Theta)} \ \ \ \ \Delta=\frac{\frac{nK NV}{3c}(N\chi_Tk_\mathrm{B}T 180\alpha G^\prime+28\beta^2_G+4\beta^2_A)}{\frac{K N V}{3}( N \chi_Tk_\mathrm{B}T90\alpha^2+14\beta^2)}. \nonumber
\end{align}
Note that the circular intensity differential $\Delta$ here reduces to the previously reported result \cite{Barron71a, Andrews80a, Barron04a, Zuber08a} when $N\chi_T k_\mathrm{B}T=1$ (as in an ideal gas). For a liquid with $N\chi_T k_\mathrm{B}T<1$, the isotropic ($\alpha^2$,$\alpha G^\prime$) contributions to $\Delta$ are suppressed relative to the anisotropic ($\beta^2$,$\beta_G^2$,$\beta_A^2$) contributions, changing the magnitude of $\Delta$ and perhaps even the sign; a demonstration of the need for our extended theory. The circular spectral differential $\bbDelta$ offers more information than $\Delta$ in that the isotropic and anisotropic contributions can be distinguished as a function of the angular frequency $\Omega$. In the limiting case where $\Gamma_\Theta\gg vq\gg\Gamma q^2>D_Tq^2$, we find in particular that
\begin{align}
\bbDelta(\omega)\approx\bbDelta(\omega\mp vq)\approx\frac{\frac{nK}{3c}180\alpha G^\prime}{\frac{K}{3}90\alpha^2} \ \ \ \ 
\lim_{|\Omega-\omega|\gg vq}\bbDelta\approx\frac{\frac{nK}{3c}(28\beta^2_G+4\beta^2_A)}{\frac{K}{3}14\beta^2}. \nonumber
\end{align}
Using typical values \cite{Zuber08a}, this leads us to predict that
\begin{align}
|\bbDelta(\omega)|\approx|\bbDelta(\omega\mp vq)|\sim 10^{-4}\textrm{–}10^{-5}\ \ \ \ 
\lim_{|\Omega-\omega|\gg vq}|\bbDelta|\sim10^{-3}\textrm{–}10^{-4}, \nonumber
\end{align}
although considerable variation is possible, of course.

These features are illustrated in Fig. \ref{Fig2} for $\omega=3.54\,\mathrm{Prad}\,\mathrm{s}^{-1}$, $n=1.40$, $\gamma=1.33$, $N\chi_T k_B T=0.100$, $D_T q^2/2\pi=80.0\,\mathrm{MHz}$, $vq/2\pi=3.00\,\mathrm{GHz}$, $\Gamma q^2/2\pi=160\,\mathrm{MHz}$, $\Gamma_\Theta/2\pi=10.0\,\mathrm{GHz}$, $\beta^2=0.100\alpha^2$, $\alpha G^\prime/c=\mp1.00\times10^{-5}\alpha^2$ and $\beta_G^2/c=\beta_A^2/c=\pm1.00\times10^{-4}\alpha^2$, where the upper and lower signs correspond to opposite enantiomers. The circular spectral sum $\mathbb{I}^{\mathrm{R}\mathrm{V}}+\mathbb{I}^{\mathrm{L}\mathrm{V}}$ (i.e. the spectrum of the $S_0$ Stokes parameter of the scattered light \cite{Barron04a}) is positive and consists of a narrow Gross (centre) line \cite{Gross30a, Landau34a} and narrow Brillouin lines \cite{Brillouin14a, Brillouin22a, Mandelstam26a, Gross30a} superposed with a broad Rayleigh wing \cite{Cabannes28a, Cabannes28b, Raman28a, Cabannes29a, Manneback30a, Manneback30b}. The circular spectral difference $\mathbb{I}^{\mathrm{R}\mathrm{V}}-\mathbb{I}^{\mathrm{L}\mathrm{V}}$ (i.e. the spectrum of the $S_3$ Stokes parameter of the scattered light \cite{Barron04a}) has opposite signs for opposite enantiomers, appearing inverted at the Gross and Brillouin lines as we have taken $180\alpha G^\prime$ and $28\beta^2_G+4\beta^2_A$ to have opposite relative signs. The resulting circular spectral differential $\bbDelta$ also has opposite signs for opposite enantiomers and appears inverted at the Gross and Brillouin lines, where it is suppressed in magnitude in accord with our prediction above.

%%%%%%%%%%%%%%%%%%%%%%%%%%%%%%%%%%%%%%
%      
%%%%%%%%%%%%%%%%%%%%%%%%%%%%%%%%%%%%%%
%%%%%%%%%%%%%%%%%%%%%%%%%%%%%%%%%%%%%%
%       CONCLUSIONS
%%%%%%%%%%%%%%%%%%%%%%%%%%%%%%%%%%%%%%
%%%%%%%%%%%%%%%%%%%%%%%%%%%%%%%%%%%%%%
%       
%%%%%%%%%%%%%%%%%%%%%%%%%%%%%%%%%%%%%%

\section{Conclusions}
\label{Conclusions}
We have presented a theory of RayBOA applicable to dense samples such as neat liquids. There are many possible avenues for future research.

We have evaluated the circular differentials for a toy model of an enantiopure neat liquid. It remains for us to consider more realistic models for a variety of different samples. Molecular dynamics simulations have recently been applied with success to ROA \cite{Brehm17a, Yang22a, Michal22a} and might be developed for RayBOA as well.

We have adopted a microscopic approach, facilitating comparison with the existing theory of RayOA \cite{Atkins69a, Barron71a, Andrews80a, Hecht94a, Craig98a, Barron04a, Zuber08a, Cameron14a, Cameron18a, Forbes19a, Forbes19b}. A macroscopic approach is also possible \cite{Smoluchowski08a, Einstein10a, Berne00a, Boyd03a, Landau84a} and might yield new insights. Care  will need to be taken with the choice of macroscopic constitutive relations \cite{Barnett16a, Ossikovski21a}, which must include electric dipole-electric quadrupole contributions to describe RayBOA correctly even for isotropic samples.

We have focussed on SCP, ICP and DCP$_\mathrm{I}$ RayBOA for off-resonant illumination of a fluid by planar light. Interesting new features should emerge for illumination near resonance \cite{Barron85a, Nafie89a, Hecht90a, Hecht94a}, anisotropic samples \cite{Hecht94a} and illumination by structured light \cite{Cameron14a, Forbes19a, Forbes19b, McArthur20a}. It should also prove fruitful to consider the influence of static magnetic fields \cite{Barron72a}, static electric fields \cite{Buckingham75a, Gasso22a} and higher-order multipolar contributions \cite{Tinoco84a, Cameron18a}.

A particularly interesting question is whether Rayleigh-Brillouin optical activity can be used to detect chiral (acoustic) phonons in appropriate samples \cite{Zhu18a, Choi22a, Michal22a, Ueda23a}.

Although our focus in this paper has been on small molecules, similar ideas can be developed for larger scatterers, including large biomolecules \cite{Maestre82a, Tinoco84a, Gratiet20a, Pan22a, Gasso22a}.

We will return to these and related tasks elsewhere.

%%%%%%%%%%%%%%%%%%%%%%%%%%%%%%%%%%%%%%
%      
%%%%%%%%%%%%%%%%%%%%%%%%%%%%%%%%%%%%%%
%%%%%%%%%%%%%%%%%%%%%%%%%%%%%%%%%%%%%%
%       ACKNOWLEDGEMENTS
%%%%%%%%%%%%%%%%%%%%%%%%%%%%%%%%%%%%%%
%%%%%%%%%%%%%%%%%%%%%%%%%%%%%%%%%%%%%%
%       
%%%%%%%%%%%%%%%%%%%%%%%%%%%%%%%%%%%%%%

\section*{Acknowledgements}
The authors gratefully acknowledge support from the Royal Society (URF$\backslash$R1$\backslash$191243) and EPSRC (EP/R513349/1). We thank Laurence Barron for useful discussions. RPC is a Royal Society University Research Fellow.

%%%%%%%%%%%%%%%%%%%%%%%%%%%%%%%%%%%%%%
%      
%%%%%%%%%%%%%%%%%%%%%%%%%%%%%%%%%%%%%%
%%%%%%%%%%%%%%%%%%%%%%%%%%%%%%%%%%%%%%
%      REFERENCES
%%%%%%%%%%%%%%%%%%%%%%%%%%%%%%%%%%%%%%
%%%%%%%%%%%%%%%%%%%%%%%%%%%%%%%%%%%%%%
%       
%%%%%%%%%%%%%%%%%%%%%%%%%%%%%%%%%%%%%%

\bibliography{Bib}

%%%%%%%%%%%%%%%%%%%%%%%%%%%%%%%%%%%%%%
%      
%%%%%%%%%%%%%%%%%%%%%%%%%%%%%%%%%%%%%%
%%%%%%%%%%%%%%%%%%%%%%%%%%%%%%%%%%%%%%
%       END OF THE DOCUMENT
%%%%%%%%%%%%%%%%%%%%%%%%%%%%%%%%%%%%%%
%%%%%%%%%%%%%%%%%%%%%%%%%%%%%%%%%%%%%%
%       
%%%%%%%%%%%%%%%%%%%%%%%%%%%%%%%%%%%%%%

\end{document}